\documentstyle[prd,aps, 
preprint,
multicol,axodraw]{revtex}

\begin{document}

\draft
\preprint{\begin{tabular}{r}
KIAS-P00005 \\
hep-ph/0001296
\end{tabular}}

\title{Baryogenesis and  Degenerate Neutrinos} 
\author{Eung Jin Chun and Sin Kyu Kang }
\address{Korea Institute for Advanced Study, 
    Seoul 130-012, Korea \\
{\it Email addresses: ejchun@kias.re.kr, skkang@kias.re.kr}    }
\maketitle

\begin{abstract}
We bring the theoretical issue of whether two important cosmological
demands, baryon asymmetry and degenerate neutrinos as hot dark matter,
can be compatible in the context of the seesaw mechanism.
To realize leptogenesis with almost degenerate Majorana neutrinos
without severe fine-tuning of parameters, we propose the hybrid seesaw
mechanism with a heavy Higgs triplet and right-handed neutrinos.
Constructing a minimal hybrid seesaw model with SO(3) flavor 
symmetry for the neutrino sector, we show that the mass splittings for the 
atmospheric and solar neutrino oscillations which are consistent with the
requirements for leptogenesis  can naturally arise.
\end{abstract}



Among various scenarios explaining the cosmological baryon asymmetry,
the most attractive one is the leptogenesis mechanism 
given the experimental indications for nonzero neutrino masses.
Current data from atmospheric \cite{skatm} and solar \cite{solex}
neutrino observations provide evidence for massive neutrinos,
and terrestrial neutrino experiments \cite{lsnd,chooz,verde} 
lead to meaningful constraints on neutrino 
masses and mixing.   
The atmospheric neutrino oscillation indicates the near maximal mixing
between $\nu_{\mu} $ and $\nu_{\tau}$ , $\sin^2 2\theta_{atm}\geq 0.85$,
with a mass squared difference
$\Delta m^2_{atm} \simeq 3\times10^{-3}~\mbox{eV}^2$\cite{rrr}.
The solar neutrino anomaly can be explained through matter enhanced
neutrino oscillation if
$3\times 10^{-6}\leq \Delta m^2_{sol} \leq 10^{-5} ~\mbox{eV}^2$ and
$2\times 10^{-3}\leq \sin^2 2 \theta _{sol}\leq 2\times 10^{-2}$
(small angle MSW), or
$10^{-5}\leq \Delta m^2_{sol} \leq 10^{-4} ~\mbox{eV}^2$,
$\sin^2 2 \theta _{sol}\geq 0.5$ (large angle MSW), $\Delta m^2_{sol}\sim
10^{-7} ~\mbox{eV}^2, \sin^2 2 \theta_{sol}\sim 1.0$ (LOW solution)
\cite{solan} and through long-distance  vacuum oscillation if
$5\times 10^{-11}\leq \Delta m^2_{sol} \leq 10^{-9} ~\mbox{eV}^2$,
$\sin^2 2 \theta _{sol}\geq 0.6$. On the other hand, the CHOOZ experiment
can constrain $\nu_e-\nu_{x}$ oscillation with $\Delta m^2_{13}\geq 10^{-3}
~\mbox{eV}^2$ \cite{chooz}, and the recent Palo Verde reactor experiment 
also indicates no observation of atmospheric $\nu_e-\nu_{x}$ oscillation
for $\Delta m^2 \geq 1.12\times 10^{-3}$ and
for $\sin^2 2\theta \geq 0.21$(for large $\Delta m^2$)~\cite{verde}.

The lightness of three active neutrinos could be a consequence of the
existence of heavy fields and lepton number violation at a high scale 
through the seesaw mechanism \cite{ss1}.  This lepton number violation
can erase the pre-existing baryon asymmetry of the universe, but can also
lead to baryogenesis above the electroweak scale.  The latter is called 
the leptogenesis mechanism  in which 
the decays of the heavy fields can generate 
a lepton asymmetry which converts into the observed baryon asymmetry 
due to the sphaleron processes \cite{fy}.
The heavy fields in the seesaw mechanism can be either the 
right-handed neutrinos \cite{ss1}  or Higgs triplets \cite{ss2}
both of which are known to yield a successful baryogenesis 
without fine-tuning of parameters \cite{lepr,lept}.   
In this scenario, the requirement for generating
the right amount of baryon asymmetry puts meaningful constraints on 
the pattern of neutrino masses and mixing \cite{conds,bane}.  

An interesting question in this regard is whether the leptogenesis
mechanism can be consistent with degenerate neutrino scenarios which
may come from another cosmological demand
for hot dark matter consisting of neutrinos \cite{chdm}.  
Taking this cosmological indication together with 
the current neutrino data coming from the atmospheric \cite{skatm},
solar neutrino \cite{solex}, the reactor \cite{chooz,verde},
and neutrinoless double-beta decay \cite{bb0n} experiments,   
one is led to a specific pattern of Majorana neutrino mass matrix 
in the leading order terms as follows \cite{degbim};
\begin{equation} \label{mn0}
 m_\nu \sim m_0 \pmatrix{ 
     0 & {1\over \sqrt{2}} & {1\over \sqrt{2}} \cr
     {1\over \sqrt{2}} & {1\over 2} &  -{1\over 2} \cr
      {1\over \sqrt{2}} & -{1\over 2} &  {1\over 2} } \,,
\end{equation}
which gives rise to three degenerate mass eigenvalues and 
bimaximal mixing for the atmospheric and solar neutrino oscillations.
Here the three neutrinos with $m_0 \approx 2$ eV can provide the right 
amount of hot dark matter to explain the structure formation 
of the universe \cite{chdm}.
Let us first recall that the degenerate mass pattern (\ref{mn0}) 
with $m_0 = {\cal O}(1)$ eV cannot yield a successful leptogenesis 
in the canonical seesaw mechanism with heavy right-handed neutrinos.  
This is because the condition for the out-of-equilibrium decay 
of a right-handed neutrino $N_1$, $\Gamma_{N_1} < H$, 
is satisfied only when \begin{equation} \label{out1r}
 m_{\nu_1} \lesssim 4\times 10^{-3} {\rm eV}
\end{equation}
for the lightest neutrino $\nu_1$ \cite{conds}.

In this letter, we investigate the possibility of realizing both
the almost degenerate neutrino mass pattern (\ref{mn0})  and 
a successful leptogenesis in the context of the seesaw mechanism.
To find a natural model of this kind, it will be important to check
whether the small mass splittings accounting for the atmospheric and 
solar neutrino oscillations are consistent with
the out-of-equilibrium conditions of the leptogenesis mechanism.
First of all, we will examine the leptogenesis in the triplet 
seesaw model recently proposed in Ref.~\cite{lept}, and show that 
the reconciliation of leptogenesis with degenerate neutrinos 
can be made with the price of fine tuning of parameters.  
Then, we will suggest the hybrid seesaw model
consisting of right-handed neutrinos and a Higgs triplet ~\cite{hybrid}, 
in which the required lepton asymmetry is generated by the decay of
heavy right-handed neutrinos, and almost degenerate neutrinos with
the desired tiny mass splittings arise in a natural manner.

\medskip

In order to realize leptogenesis in the seesaw mechanism with 
heavy Higgs triplets, one needs at least two Higgs triplets  \cite{lept}.  
In this model, the couplings of the heavy Higgs triplets $\Delta_i$ 
with the lepton doublets $L_\alpha$ and Higgs doublet $H$ are given by
\begin{equation}
 {\cal L} = {1\over2}h_{i\alpha\beta} L_\alpha L_\beta \Delta_i
          + \mu_i H H \Delta_i + \cdots  \,.
\end{equation}
Here we take $\mu_i \sim M_i$ where 
$M_i$ is the mass of the Higgs triplet $\Delta_i$.  
Neutrino masses, then,  come from the nonvanishing vacuum expectation 
values of the neutral components of the Higgs triplets and 
the resulting neutrino mass matrix is 
\begin{equation} \label{mnt}
 m_{\nu\alpha\beta} = m_{1\alpha\beta} + m_{2\alpha\beta}  \equiv
 h_{1\alpha\beta}{\mu_1 v^2 \over M_1^2} +
 h_{2\alpha\beta}{\mu_2 v^2 \over M_2^2} 
\end{equation}
where $v \equiv \langle H \rangle$ and the mass mixing between $\Delta_1$ 
and $\Delta_2$ is neglected.  
A key ingredient for the lepton asymmetry is the one-loop CP-violating
mass correction in the decay of the lighter Higgs triplet, say, $\Delta_1$,
and it can be written as
\begin{equation} \label{epL2}
 \varepsilon_L \approx {1\over 8\pi} \sum_{\alpha\beta}
    { m_{1\alpha\beta} m_{2\alpha\beta} \over m^2_{\nu\alpha\beta}}
    {   m^2_{\nu\alpha\beta} M_1^2 M_2^2 \over v^4 (M_1^2-M_2^2)}
    {1 \over \sum_{\gamma\delta}|h_{1\gamma\delta}|^2 }
\end{equation}
where we take the CP phase of order 1 from the result of Ref.~\cite{lept}.

The quantity (\ref{epL2}) is constrained by the out-of-equilibrium 
conditions for the baryogenesis.  First, 
let us recall  that the effective operator $(m_{\nu\alpha\beta}/v^2)
L_\alpha L_\beta \bar{H} \bar{H}$ generated below the scale $M_1$ or $M_2$
should be out-of-equilibrium  in order not to erase 
the lepton asymmetry generated at the temperature $T_{B-L}$.
This gives rise to \cite{bane}
\begin{equation} \label{out1}
  T_{B-L}  \lesssim 10^{11} 
  \left(m_{\nu\alpha\beta} \over 1 {\rm eV}\right)^2 {\rm GeV}\,,
\end{equation}
where $T_{B-L}=M_1$ in our scenario under the consideration.
Second, the decay of $\Delta_1$ should also be out-of-equilibrium, 
$\Gamma_{\Delta_1} < H$, leading to  
$ \sum_{\gamma\delta} |h_{1\gamma\delta}|^2 M_1/8\pi 
 <  1.7 \sqrt{g_*} {T^2 / M_{Pl}}$ 
 at the temperature $T=M_1$.  With $g_* \sim 100$, we then have
\begin{equation} \label{out2}
   \sum_{\gamma\delta} |h_{1\gamma\delta}|^2 < 10^{-6}  
    \left(M_1 \over 10^{11} {\rm GeV} \right)\,.
\end{equation}
Since the baryon asymmetry is related to the lepton asymmetry by
$n_B/s \approx \kappa \varepsilon_L/g_* \approx 10^{-10}$,
we can estimate
$\varepsilon_L \approx 10^{-5}-10^{-7}$
for $\kappa \approx 10^{-1}-10^{-3}$ and $g_*\sim 100$. 
Combining this with the 
out-of-equilibrium conditions (\ref{out1}, \ref{out2})
and assuming that there are no fine cancellations in Eq.~(\ref{epL2}), 
we get
\begin{equation} \label{mcond}
 {m_{1\alpha\beta} m_{2\alpha\beta} \over m^2_{\nu\alpha\beta} }
 \left( {m_{\nu\alpha\beta} \over 1 {\rm eV}}\right) 
 \lesssim 10^{-6}  \,.
\end{equation}
This result implies a large hierarchy between $m_{1\alpha\beta}$ and 
$m_{2\alpha\beta}$, in other words,
$m_{1\alpha\beta} \ll m_{2\alpha\beta} \sim m_{\nu\alpha\beta}$ with 
$m_{1\alpha\beta}/m_{2\alpha\beta} \lesssim 10^{-6}$.
To achieve such a large hierarchy, an unpleasant fine tuning between
parameters related to two Higgs triplets is needed;
\begin{equation} \label{hcond}
 {h_{1\alpha\beta}\over h_{2\alpha\beta}} {\mu_1 \over \mu_2} 
 \lesssim 10^{-6} \left(M_1\over M_2\right)^2 \,.
\end{equation}
Note that we have the condition $M_1 < M_2$.

\medskip

At this point, we pay attention to another important theoretical issue 
concerned with degenerate neutrinos.  
For the Majorana neutrino mass matrix (\ref{mn0}) to be realistic, 
it has to be completed with the next leading terms which lift the degeneracy
by the small amounts so as to accommodate the atmospheric and solar neutrino 
observations, simultaneously.  Defining the quantities $\epsilon_a \equiv
(m_{\nu_3}-m_{\nu_2})/m_0$ and $\epsilon_s \equiv (m_{\nu_2}-m_{\nu_1})/m_0$
with the mass eigenvalues, $m_{\nu_3} \gtrsim m_{\nu_2} \gtrsim m_{\nu_1}$, 
the observed  mass-squared differences, $\Delta m^2_{atm}$ and 
$\Delta m^2_{sol}$, respectively for the atmospheric and solar neutrino 
oscillations fix their values as
\begin{equation} \label{epas}
 \epsilon_a = {\Delta m^2_{atm} \over 2m_0^2} \quad \mbox{and} \quad
 \epsilon_s= {\Delta m^2_{sol} \over  2m_0^2}   \,.
\end{equation}
Therefore, we have  $\epsilon_a \sim 10^{-3}$ for the atmospheric
oscillation \cite{rrr} and 
$\epsilon_s \sim 10^{-5}$, $10^{-7}$ or $10^{-10}$ 
for the large mixing angle MSW solution (LMA), 
the low $\Delta m^2$ MSW solution (LOW)
or the vacuum oscillation solution (VO) to the solar neutrino problem, 
respectively \cite{solan}.

Given the two contributions to the neutrino mass matrix (\ref{mnt}),
an interesting question one can address is whether one contribution 
corresponds to a large mass of order of $m_0$ and the other to 
the tiny splitting $\epsilon_a$ or $\epsilon_s$.
The required hierarchy (\ref{mcond}) shows that the ratio 
$m_{1\alpha\beta}/m_{2\alpha\beta}$ cannot give rise to $\epsilon_a$, 
but it can be used to generate $\epsilon_s$ for the case of the LOW or 
VO solution.  
That is, the splitting $\epsilon_a$ and $\epsilon_s$ 
for the LMA solution have to be arranged 
within the mass matrix $m_2$ in the triplet seesaw model. 

\medskip

To remedy the fine-tuning problem in realizing both the leptogenesis 
mechanism and three degenerate neutrinos in the triplet 
seesaw mechanism, let us suggest a simple hybrid model 
with a Higgs triplet and three heavy right-handed neutrinos 
which allows for the Yukawa and 
Higgs couplings 
\begin{equation} \label{Lhyb}
 {\cal L}= {1\over2} h_{\alpha\beta}L_\alpha L_\beta\Delta
    + f_{\alpha\beta}L_\alpha N_\beta \bar{H}+\mu H H \Delta +\cdots \,.
\end{equation}
Let $M_\Delta$ and $M_N$ are the masses of the Higgs triplet and the
right-handed neutrinos, respectively, and $\mu$ parameter is taken
to be of order $M_{\Delta}$.  We will further assume that 
$M_\Delta \gtrsim M_N$ and the lepton asymmetry arises from the decay of
the heavy right-handed neutrinos.   
If we take the decay of the heavy Higgs triplet as the origin 
of the lepton asymmetry, we encounter the similar fine-tuning problem
as in Eq.~(\ref{hcond}).
There are again two contributions to the neutrino mass matrix given by
\begin{equation} \label{mnhyb}
 m_\nu = m_1 + m_2 \equiv 
  {f^2 v^2 \over M_N} + h{\mu v^2\over M_\Delta^2},,
\end{equation}
where we neglected  the flavor indices of the parameters $f, h$ and $M_N$.  
Now, the out-of-equilibrium
conditions are satisfied when
\begin{eqnarray} \label{out1h}
 m_1 &\lesssim& 4\times10^{-3} {\rm eV}\,,  \\
 \label{out2h}
 M_N &\lesssim& 10^{11} 
  \left(m_{\nu\alpha\beta} \over 1 {\rm eV}\right)^2 {\rm GeV}\,,
\end{eqnarray}
which are the counterparts of the previous Eqs.~(\ref{out1r}) 
and (\ref{out1}), respectively.   
The condition (\ref{out1h}) implies the hierarchy
$m_1 \ll m_2 \sim 1$ eV with $m_1/m_2 \lesssim 10^{-3}$, which would be
relevant for the required splitting $\epsilon_a$ for the atmospheric
neutrino oscillation.  The CP-nonconservation in our hybrid model 
is generated by the interference
between the tree and one-loop diagram mediated by the Higgs triplet 
as shown in Fig.~1, and the resulting lepton asymmetry is given by
\begin{equation} \label{epLh}
 \varepsilon_L \approx {1\over 8\pi} 
 { {\rm Im}(f^2h^*\mu) \over M_N |f|^2 }F({M_\Delta^2\over M_N^2}) 
\end{equation}
where $F(x)=\sqrt{x}[1-(1-x)\ln(1+x)/x]$.
Taking the CP phase of order 1, we thus have
\begin{equation}
 \varepsilon_L \approx 10^{-5} \left( m_2 \over 1{\rm eV}\right)
 \left( M_\Delta \over 10^{10} {\rm GeV} \right)
 \left( M_\Delta \over M_N \right) \,,
\end{equation}
which provides the right amount of the lepton asymmetry for
$M_N \sim M_\Delta \sim (10^8-10^{10})$  GeV
while satisfying the out-of-equilibrium condition (\ref{out2h}).

As alluded above, it is amusing to observe that the splitting 
$\epsilon_a$ can be provided by the ratio $m_1/m_2$ satisfying 
the leptogenesis 
requirements, contrary to the triplet seesaw model.
Requiring $m_1\sim 10^{-3}$ eV and $m_2 \sim 1$ eV, therefore,
we find
\begin{eqnarray} \label{fh}
 f\sim 4\times 10^{-4} \left(M_N \over 10^{10} {\rm GeV}\right)^{1/2}\,,
            \nonumber\\
 h\sim 10^{-4} \left(M_\Delta \over \mu\right)
 \left(M_\Delta \over 10^{10} {\rm GeV}\right) \,.
\end{eqnarray}
Thus, we can accomplish leptogenesis in the hybrid model with three
almost degenerate light neutrinos when both Yukawa couplings $f$ and $h$ 
are of order $10^{-4}$.   Therefore, we one can avoid a big hierarchy 
between the parameters of the theory.  Still, it remains to be understood
the overall smallness of our parameters; $f\sim h \sim 10^{-4}$ and 
$\mu \sim M_N \sim M_\Delta \sim 10^{-10}$ GeV, which would be resolved
with the question of an intermediate scale.
Having $\epsilon_a \sim m_1/m_2$, let us remark that 
the splitting $\epsilon_s$ can come from the $\tau$ Yukawa coupling 
effect through the renormalization group evolution \cite{rge,rge2}. 
It is then enough to introduce only two Yukawa couplings 
for our purpose: one $h$ for generating $m_0$ and one $f$ for 
$\epsilon_a$.

\medskip

{}From now on, we construct the minimal hybrid seesaw model
accommodating all the features under consideration.
For this, we rely on 
the symmetry principle from which the degenerate mass $m_0$ 
and the relevant splittings $\epsilon_a$ and $\epsilon_s$ 
are obtained in a systematic way.  Let us consider 
the SO(3) flavor symmetry under which 
the lepton doublets form a triplet with the $(+,-,0)$ components.
The SO(3) symmetry has to be badly broken by the charged-lepton Yukawa
couplings and the working hypothesis is that the SO(3) flavor basis is
related to the charged-lepton flavor basis as follows \cite{ma2};
\begin{equation} \label{U0}
 L_+ = L_e,\; L_-=c_1 L_\mu- s_1 L_\tau,\;
 L_0=s_1 L_\mu+ c_1 L_\tau 
\end{equation}
where $c_1=\cos\theta_1$, {\it etc}. Recall that the angle $\theta_1$ 
is determined by the atmospheric neutrino mixing, that is, 
$c_1^2\sim s_1^2 \sim 1/2$.
We further assume that the lepton doublet couplings with the Higgs 
triplet $\Delta$ and a right-handed neutrino preserve the SO(3) 
symmetry and its U(1) subgroup, respectively as follows;
\begin{equation} \label{mhyb}
 {\cal L} = h(L_+ L_- +{1\over2} L_0L_0) \Delta + fL_0 N_0 \bar{H} 
          + \mu H H \Delta + \cdots
\end{equation}
where we introduced only one right-handed neutrino $N_0$ with the U(1) charge 
0 as a minimal choice.  Our conclusion is not altered by introducing three
right-handed neutrinos as long as their couplings and mass terms preserve
the U(1) subgroup \cite{kang}.  The full neutrino mass matrix gets important 
contributions not only from the Lagrangian (\ref{mhyb}) at tree level
but also from the one-loop correction due to the tau Yukawa coupling $h_\tau$
\cite{rge,rge2}.  The latter contribution breaks the U(1) subgroup of 
the SO(3) flavor symmetry and is controlled by the quantity 
$\epsilon_\tau \equiv h_\tau^2 \ln(M_N/M_Z)/32\pi^2 \approx 10^{-5}$.
Including all these contributions, we get the neutrino mass matrix
in the SO(3) basis;
\begin{equation} \label{mnmh}
 m_\nu = m_0 \pmatrix{ 0 & (1+{1\over2}s_1^2 \epsilon_\tau) & 
 -{1\over2}c_1s_1\epsilon_\tau  \cr 
 (1+{1\over2}s_1^2 \epsilon_\tau)  & 0 & 
  -{1\over2}c_1s_1\epsilon_\tau  \cr 
  -{1\over2}c_1s_1\epsilon_\tau   &  -{1\over2}c_1s_1\epsilon_\tau  & 
  (1+c_1^2\epsilon_\tau)_0 + {\delta m_0 \over m_0} \cr}
\end{equation}
where $m_0 \equiv  h \mu v^2/M_\Delta^2$ and 
$\delta m_0 \equiv  f^2v^2/M_N$ as in Eq.~(\ref{mnhyb}). 
Transformed into the charged-lepton flavor basis by Eq.~(\ref{U0}),
the leading terms of the matrix (\ref{mnmh}) reproduce the desired form of
mass matrix (\ref{mn0}).  From Eq.~(\ref{mnmh}), 
the quantities $\epsilon_a$ and $\epsilon_s$ can be calculated as
\begin{equation}
 \epsilon_a \approx {\delta m_0 \over m_0 } \,,\quad 
 \epsilon_s \approx {1\over4}\sin^22\theta_1 
              {\epsilon_\tau^2 \over \epsilon_a} \,.
\end{equation}
With $\epsilon_a \sim 10^{-3}$ required by the atmospheric neutrino
oscillation, we have $\epsilon_s \sim 10^{-7}$ which is in the right range
for the LOW solution.    As noted in Ref.~\cite{kang}, the LMA solution
(requiring $\epsilon_s \sim 10^{-5}$)
can be realized in the two Higgs doublet model where $\epsilon_s$ contains
the additional factor $\tan^4\beta$.  For the successive
SO(3) breaking to be realistic,  the hierarchy between the Yukawa couplings,
$f\ll h$, would have to be imposed as the latter conserves the SO(3)
flavor symmetry and the former breaks it.  This would require 
$\mu/M_\Delta \approx 0.1-0.01$ as can be seen from Eq.~(\ref{fh}). 

\medskip

In conclusion, we have brought the theoretical issue of reconciling 
two important cosmological demands, baryon asymmetry and 
neutrino as hot dark matter, in the context of the seesaw mechanism.
For this purpose, we have examined whether the almost degenerate 
mass pattern accounting for hot dark matter and the other neutrino data 
can be consistent with the leptogenesis mechanism in 
various types of the seesaw models.  
As was pointed out, this feature cannot be realized in the canonical
seesaw mechanism. On the other hand, we have shown that some fine-tuning 
between the couplings related to each Higgs triplet is needed
to achieve leptogenesis with almost degenerate neutrinos 
in the triplet seesaw model.  To resolve this problem, 
we have suggested the hybrid seesaw mechanism with 
a heavy Higgs triplet and right-handed neutrinos.
In this type of models, the out-of-equilibrium conditions required 
for the successful baryogenesis can be naturally satisfied with
the almost degenerate neutrino masses and the small mass 
splitting for the atmospheric neutrino oscillation.  
Furthermore, the mass splitting for the solar
neutrino can come from the renormalization group effect due to the 
tau Yukawa coupling.  Finally, we have presented a simple model realizing
all these features.  In the minimal case, the model consists of a
Higgs triplet and a right-handed neutrino with two additional leptonic 
Yukawa couplings.  These two parameters are responsible for generating 
the degenerate neutrino mass in the leading order and the splitting 
for the atmospheric neutrino mass-squared difference, which respect
the SO(3) flavor symmetry and its U(1) subgroup, respectively.


\bigskip
\begin{figure}
\begin{center}
\begin{picture}(300,150)(0,0)
\Vertex(50,60){2}
\ArrowLine(0,60)(50,60)
\ArrowLine(50,60)(90,30)
\DashLine(50,60)(90,90){2}
\Text(25,70)[]{$N$}
\Text(65,35)[]{$L$}
\Text(65,85)[]{$H$}
\Text(125,60)[]{$+$}
\Vertex(200,60){2}
\Vertex(240,90){2}
\Vertex(240,30){2}
\ArrowLine(150,60)(200,60)
\ArrowLine(200,60)(240,30)
\ArrowLine(240,30)(280,0)
\DashLine(200,60)(280,120){2}
\DashLine(240,30)(240,90){2}
\Text(175,70)[]{$N$}
\Text(215,40)[]{$L$}
\Text(215,80)[]{$H$}
\Text(250,60)[]{$\Delta$}
\Text(260,5)[]{$L$}
\Text(260,115)[]{$H$}
\end{picture} 

Figure~1. The tree and one-loop diagrams generating the lepton asymmetry.
\end{center}
\end{figure}

\begin{thebibliography}{99}
\def\plb#1#2#3{Phys.\ Lett.\       {\bf B#1}, #2 (#3)}
\def\npb#1#2#3{Nucl.\ Phys.\       {\bf B#1}, #2 (#3)}
\def\prd#1#2#3{Phys.\ Rev.\        {\bf D#1}, #2 (#3)}
\def\prl#1#2#3{Phys.\ Rev.\ Lett.\ {\bf #1},  #2 (#3)}
\def\mpl#1#2#3{Mod.\ Phys.\ Lett.\ {\bf A#1}, #2 (#3)}
\def\rep#1#2#3{Phys.\ Rep.\        {\bf #1},  #2 (#3)}
\def\sci#1#2#3{Science             {\bf #1},  #2 (#3)}
\def\astro#1#2#3{Astrophys.\ J.\   {\bf #1},  #2 (#3)}
\bibitem{skatm}  
    The Super-Kamiokande Collaboration, Y. Fukuda, {\it et.al.},
    \prl{81}{1562}{1998}.
\bibitem{solex}
 B. T. Cleveland {\it et al.}, Astrophys. J. {\bf 496}, 505 (1998); 
 Kamiokande Collaboration, K. S. Hirata {\it et al.}, \prl{77}{1683}{1996};
 GALLEX Collaboration, W. Hampel {\it et al.}, \plb{447}{127}{1999}; 
 SAGE Collaboration, J. N. Abdurashitov {\it et al.}, astro-ph/9907113; 
 Super-Kamiokande Collaboration, Y. Fukuda, {\it et al.}, \prl{82}{2430}{1999}.

\bibitem{lsnd} 
  The LSND Collaboration, C. Athanassopoulos {\it et al.},
  \prl{81}{1774}{1998}.
\bibitem{chooz} The CHOOZ Collaboration, M. Apollonio {\it et al.},
         \plb{420}{397}{1998}; hep-ex/9907037.
\bibitem{verde} The Palo Verde Collaboration, F. Boehm {\it et al.}, 
         hep-ex/9912050; hep-ex/0003022.
\bibitem{rrr} For a recent global analysis, wee N.Forrengo, 
M.C. Gonzales-Garcia, and J.W.F. Valle, hep-ph/0002147.
\bibitem{solan}  For recent analyses, see
    J. N. Bahcall, P. I. Krastev and A. Yu. Smirnov, 
    \prd{58}{096016}{1998};
    M.C. Gonzalez-Garcia, P.C. de Holanda, C. Pena-Garaya nd J.W.F. Valle,
    hep-ph/9906469;
    G.L. Fogli, E. Lisi, D. Montanino and A. Palazzo, hep-ph/9912231.
\bibitem{ss1} 
   M. Gell-Mann, P. Ramond and R. Slansky, in 
   {\it Supergravity}, {\it ed.}\ P. van Nieuvanhuizen and D.Z. Freedman
   (North-Holland, Amsterdam, 1979);
   T. Yanagida, KEK Report No.\ 79-18 (1979).
\bibitem{fy}
   M. Fukugita and T. Yanagida, \plb{174}{45}{1986};
   M.A. Luty, \prd{45}{415}{1992}.
\bibitem{ss2} 
    R.N. Mohapatra and G. Senjanovic, \prd{23}{165}{1981}; 
    C. Wetterich, \npb{187}{343}{1981}. 
\bibitem{lepr}
   M. Flanz, E.A. Psachos and U. Sarkar, \plb{345}{248}{1995};
   L. Covi, E. Roulet and F. Vissani, \plb{384}{169}{1996};
   W. Buchm\"uller and M. Pl\"umacher, \plb{431}{354}{1998};
   A. Pilaftsis, Int. J. Mod. Phys. {\bf A14}, 1811  (1999).
\bibitem{lept}
    E. Ma and U. Sarkar, \prl{80}{5716}{1998}.
\bibitem{conds}
   W. Fischler, G.F. Giudice, R.G. Leigh and S. Paban, \plb{258}{45}{1991};
   W. Buchm\"uller and T. Yanagida, \plb{302}{240}{1993}.
\bibitem{bane}
  J.A. Harvey and M.S. Turner, \prd{42}{3344}{1990};
  A.E. Nelson and S.M. Barr, \plb{246}{141}{1990}.
\bibitem{chdm}
  E. Gawiser and J. Silk, Science {\bf 280}, 1405 (1998);
  J.R. Primack and M.A.K. Gross, hep-ph/9810204.
\bibitem{bb0n} 
  L. Baudis {\it et al.}, hep-ex/9902214.
\bibitem{degbim} 
  F.Vissani, hep-ph/9708483; 
  V. Barger, S. Pakvasa, T.J. Weiler and K. Whisnant, \plb{437}{107}{1998};
  H. Georgi and S.L. Glashow, hep-ph/9808293.
\bibitem{hybrid}
 For earlier attempts to get dengenerate neutrinos, see
    S.T. Petcov and A. Yu Smirnov, \plb{322}{109}{1994};
    A.S. Joshipura, Z.\ Phys.\ C64, 81 (1994);
    P. Bamert and C.P. Burgess, \plb{329}{289}{1994}
    A. Ioannissyan and J. W. F. Valle, \plb{332}{93}{1994};
    D.G. Lee and R.N. Mohapatra, \plb{329}{463}{1994};
    R. N. Mohapatra and S. Nussinov, \plb{346}{75}{1995}.
\bibitem{rge} 
    P.H. Chankowski and Z. Pluciennik, \plb{316}{312}{1993};
    K. Babu, C. N. Leung and J. Pantaleone, \plb{319}{191}{1993}.
    M. Tanimoto, \plb{360}{41}{1995}.
\bibitem{rge2}
    J. Ellis and S. Lola, \plb{458}{310}{1999}; 
    N. Haba {\it et al}, Eur.\ Phys.\ J. {\bf C10}, 677 (1999); 
    hep-ph/9906481;  hep-ph/9911481; 
    J. A. Casas, J. R.  Espinosa, A. Ibarra and I. Navarro, 
    \npb{556}{3}{1999}; JHEP {\bf 09}, 015 (1999); hep-ph/9910420;
    R. Barbieri, G. G. Ross and A. Strumia, JHEP 9910:020 (1999);
    P.H. Chankowski, W. Krolikowski and S. Pokorski, 
    \plb{473}{109}{2000};
    E.J. Chun and S. Pokorski, hep-ph/9912210.
\bibitem{ma2} 
   E. Ma, \plb{456}{48}{1999}; \prl{83}{2514}{1999}.
\bibitem{kang}
  E.J. Chun and S.K. Kang, hep-ph/9912524.

\end{thebibliography}
\end{document}